\documentclass[prd, floatfix,twocolumn,nofootinbib]{revtex4}
\usepackage{hyperref}
\usepackage{amssymb}
\usepackage{amsmath}

\usepackage{epstopdf}
\usepackage{subfigure}

\usepackage{graphicx}

\def\ba{\begin{eqnarray}}
\def\ea{\end{eqnarray}}
\def\be{\begin{equation}}
\def\ee{\end{equation}}
\def\ben{\begin{equation} \nonumber}
\def\een{\end{equation}}
\def\lse{{\,\overset{<}{_\sim}\,}}

\def\lb{\left(}
\def\rb{\right)}
\def\ls{\left[}
\def\rs{\right]}
\def\lc{\left\{}
\def\nn{\nonumber}
\def\rc{\right\}}

\begin{document}
\bibliographystyle{prsty}
\title{Correlations between 21~cm Radiation and the CMB from Active Sources}
\author{Aaron Berndsen$^{1,2}$, Levon Pogosian$^{1,3}$ and Mark Wyman$^3$}
\affiliation{$^1$Department of Physics, Simon Fraser University, Burnaby, BC, V5A 1S6, Canada. \\ 
 $^2$Department of Physics \& Astronomy, University of British Columbia, Vancouver, BC, V6T 1Z1, Canada \\
$^3$Perimeter Institute for Theoretical Physics, Waterloo, ON, L6H 2A4, Canada. } 

\date{\today}

\begin{abstract}
Neutral hydrogen is ubiquitous, absorbing and emitting 21~cm radiation
throughout much of the Universe's history. Active sources of
perturbations, such as cosmic strings, would generate simultaneous
perturbations in the distribution of neutral hydrogen and in the Cosmic
Microwave Background (CMB) radiation from recombination. Moving strings
would create wakes leading to 21~cm brightness fluctuations, while also
perturbing CMB light via the Gott-Kaiser-Stebbins effect. This would
lead to spatial correlations between the 21~cm and CMB
anisotropies. Passive sources, like inflationary perturbations, predict
no cross correlations prior to the onset of reionization. Thus,
observation of any cross correlation between CMB and 21~cm radiation
from dark ages would constitute evidence for new physics. We calculate
the cosmic string induced correlations between CMB and 21~cm and
evaluate their observability. 
\end{abstract}

\maketitle

%begin 
\section{Introduction}
\label{sec:intro}
Neutral hydrogen emits and absorbs radiation at 21~cm in its rest frame
throughout the history of the Universe. The brightness of the 21~cm line
observed today from various redshifts is determined by the spatial
distribution and peculiar velocity of the neutral hydrogen that emitted
it. The radiation from any particular redshift has small angular
variations, similar to those in the Cosmic Microwave Background (CMB)
radiation. There is also variation along the line of sight, with a
correlation depth of approximately
$10$~Mpc~\cite{Eisenstein:2005su,Furlanetto:2006jb}. In the absence of
physics that violates free streaming, one does not expect a correlation
among slices in the line-of-sight direction that are separated by more
than this correlation length. Since observations can measure the
redshift of 21~cm emission precisely, we should be able to extract
information about this line of sight variation from 21~cm observations.  

It is this third dimension of information encoded in 21~cm radiation
that can make it a powerful new tool to learn about fundamental
physics. One path forward is to look for physics that violates free
streaming and generates correlations among distinct redshift
shells~\cite{Furlanetto:2006jb}. Several sources of such a signal have
been identified, though they are limited to the epoch during and after
reionization~\cite{Alvarez:2005sa,21cm-xcorr,Adshead:2007ij,Sarkar:2008sz}. Though
passive perturbations, such as those seeded by inflation, cannot
generate such correlations, active sources of perturbation could. Active
sources of perturbation include cosmic strings and other topological
defects, such as textures \cite{Turok:1989ai}, as well as the field
gradients that are left behind after global phase transitions
\cite{JonesSmith:2007ne}. Of these, cosmic strings are the most
thoroughly studied and have perhaps the richest phenomenology
\cite{Myers:1900zz}, so they will be our focus.
 
Moving strings generate density wakes in the material through which they
pass \cite{wakes}, including neutral hydrogen. As a result, the
brightness of the 21~cm emission, which depends on the density and the
velocity of the emitters, will be directly perturbed in the string's
wake~\cite{Khatri:2008zw}. The same moving strings also perturb the
background CMB light, monotonically shifting the spectra and creating
line discontinuities in the temperature screen of the sky---a phenomenon
known as the Gott-Kaiser-Stebbins (GKS) \cite{gott,kaiser} effect.  GKS
is a special case of the Integrated Sachs-Wolfe (ISW) effect
\cite{Sachs:1967er} caused by time-varying gravitational potentials
sourced by moving strings. Thus, there will be some correlation between
string-sourced 21~cm brightness fluctuations at any redshift and a part
of the CMB temperature anisotropy. Strings will also induce some
cross-correlation in 21~cm fluctuations coming from different
redshifts. However, as we discuss in Section~\ref{sec:physics}, we
expect this effect to be relatively small.  

Precision measurements of the CMB temperature anisotropy limit the
contribution of cosmic string perturbations to be less than about 10\%
of the total over the range of scales covered by
WMAP~\cite{Wyman:2005tu}. As such, the bulk of the Universe's anisotropy
comes from the primordial inflationary fluctuations, for which the
cross-correlation between sufficiently-separated shells is zero. The
fact that any string contribution to primordial anisotropy must be small
means that the intrinsic signal for which we are searching will be
faint. However, since there is nothing in the standard $\Lambda$CDM
model that could generate such cross correlations before the epoch of
reionization, we can be sure that discovery of such a cross correlation
would be a sign of new physics.

To examine the prospects for detecting these string-induced correlations,
we will evaluate the signal-to-noise density of this effect and the
volume over which the signal can be found. Our aim is to identify the
optimal way of looking for stringy phenomena in the 21~cm radiation from
the dark ages. This effect is present over a vast volume---from
$z\lse100$ until the onset of reionization---which may make a detection
of the CMB/21cm correlation feasible under optimistic observing
scenarios. However, because of uncertainties in reionization physics, we
exclude a large region ($z<20$) from our present study. We will also
leave the calculation of  21cm $\times$ 21cm cross-correlation from
different redshifts for future work.  

In this paper we adopt the string network model of
\cite{vectorstring,ABR97}, which was previously implemented in CMBFAST
\cite{cmbfast} and made available publicly as CMBACT~\cite{cmbact}. For
this work, we have implemented this string model in
CAMB~\cite{Lewis:2007kz}. Our code will be made publicly available soon
\cite{ustobe}.

\section{21~cm/CMB correlation induced by cosmic strings}
\label{sec:physics}

Cosmic strings are relativistically moving linear topological defects
that are remnants of the high energy early Universe. As they move,
strings affect matter by kicking it into over-dense wakes and perturb
light by shifting its spectrum via the GKS effect. Strings form in the
early Universe and remain active through all epochs, spreading their
effects throughout the Universe's history. 

The effect of cosmic string networks on the CMB power spectrum has been
well studied. Strings affect the CMB in both of the ways described
above: their wakes  generate a single peak at an angular scale
determined by the strings' correlation length ($\lesssim c/H$), and
strings in the foreground of the CMB generate an interlacing network of
linear discontinuities through the GKS effect. The fact that these
signatures have not been observed provide a bound on the density and
tension of cosmic strings \cite{Wyman:2005tu}.  

These effects are also present during the dark ages, when the neutral
hydrogen is releasing 21~cm radiation. Moving strings produce wakes and
thus seed hydrogen density contrasts that grow and generate observable
variations in the brightness temperature. Interestingly, though, the GKS
effect is much less observable for 21~cm radiation. Strings shift the
spectrum of light that passes them, so their effect on passing light is
only observable when there is a noticeable change in the known
background source, such as a shift in a spectral peak or discontinuity
in a smoothly varying brightness. However, in the presence of a flat
spectrum of background light, the string is invisible, since a lateral
shift of a horizontal spectrum results in no detectable change. In the
dark ages, strings will be swimming in a background of continuously
renewed radio waves, with a very weak gradient in brightness and
spectrum. Thus, the strings' GKS effect on 21~cm radiation will be
considerably less observable than the same effect applied to CMB
photons\footnote{We thank Mark Halpern for a discussion of this
effect.}. 

To calculate these string-sourced effects quantitatively, we must give a
formal characterization of the string perturbations. The most important
difference between string-sourced perturbations and those from
inflation, besides the active/passive distinction, is that strings
source vector mode perturbations. For strings, vector modes are as large
as scalar modes, and they also generate much of the novel phenomenology
of string networks \cite{vectors}. Vector modes are usually neglected
since they decay in systems described by General Relativity and ordinary
matter. Strings, however, actively source vector perturbations faster
than they can decay away.  We will reproduce the most relevant equations
for vector mode perturbations in the Appendix. 

Calculating CMB and 21~cm observables also requires a string model and a
code for tracing the history of light streaming to us now from these
early times. To this end, we have incorporated a model for strings
\cite{ABR97,vectorstring,cmbact} (see \S~\ref{model} for a description) into
the 21~cm extension of CAMB~\cite{Lewis:2007kz}; this code will be
publicly released \cite{ustobe}. Our major addition to the underlying
CAMB code is the inclusion of vector mode fluctuations. Details about
how string-sourced vector modes enter the Boltzmann equation can be
found in Refs.~\cite{ABR97,vectorstring, Hu:1997hp, Lewis:2007kz}, and
will be more comprehensively explained in \cite{ustobe}.   

Now we can ask what physics will control the shape and amplitude of
string-sourced two-point functions in the 21~cm fluctuations and in the
cross correlation. The two physical effects controlling the shape of
these spectra are 1) the intrinsic brightness of the 21~cm photons and
2) the correlation length of the string network. The brightness
temperature sets the amplitude of the spectrum. It, in turn, is
determined by the difference between the 21~cm spin temperature and the
CMB temperature at that epoch, which are mismatched over a long period
during the dark ages \cite{Furlanetto:2006jb}. It is worth noting that
the brightness temperature is negative throughout most of the dark
ages. That is, the neutral hydrogen is absorbing, not emitting,
radiation. This shows up as an observable because we can infer what the
flux of long wavelength CMB photons should be, and hence detect the
dimming due to neutral hydrogen.  Since we are looking for small
variations in the brightness temperature, a large intrinsic brightness
makes detections easier. The anisotropy of 21~cm brightness is chiefly
caused by variations in the density and the peculiar velocity of the
neutral hydrogen, both of which are sourced by cosmic strings. These
effects and others are more comprehensively discussed in
Ref.~\cite{Lewis:2007kz}.

\begin{figure}[htbp]
\includegraphics[width=.48\textwidth]{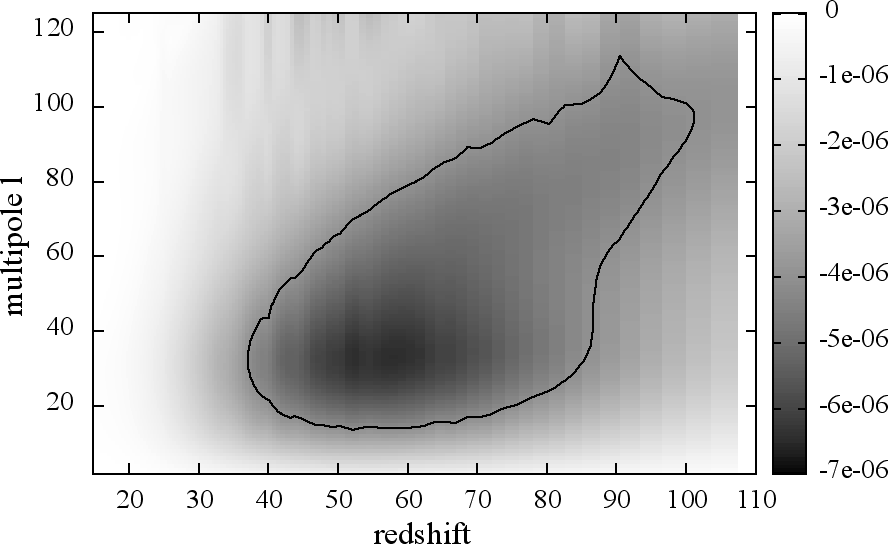}
\caption{CMB-21~cm cross-correlation power spectra 
($C_\ell^{Tz}  [\mu K]^2$) generated by a network of strings with
 tension $G\mu=4\times10^{-7}$. The amplitude of the signal scales
  with the brightness temperature, peaking at a redshift
  $z\simeq55$. The general shape---anti-correlation with a single
 broad peak---is similar for all redshifts, though the $\ell$ of
 maximum correlation moves to larger scales at later times / lower
 redshifts. This pattern comes from the string network's scaling
 behaviour: strings seed fluctuations at a fixed fraction of the horizon
 size at the epoch in which the seed is created. This corresponds to
 larger scales (smaller $\ell$'s) for later times, i.e., for lower
 redshifts. A contour at $-4\times10^{-6} [\mu K]^2$ is shown.} 
  \label{fig:Tz}
\end{figure}

The correlation length of the string network sets the location of the
peak in $\ell$-space, the angular scale. Since this correlation length
is some fraction of the Hubble scale, the spectrum from dark ages
strings peaks for multipoles $\ell<100$. This peak in the two-point
function corresponds to the size of the wakes generated by the string
network at that epoch. Hence, the peak signal migrates to lower-$\ell$
as the redshift decreases since the network is scaling with the horizon,
and the horizon is growing. This trend can be seen in our
Fig.~\ref{fig:Tz}, a plot of the CMB-21~cm cross-correlation
$C_\ell^{Tz}(z,\ell)$ from a string network, which we have generated
with our code.

\subsection{The string model}
\label{model}

The string code we used is based on the segment model
\cite{ABR97,vectorstring}. In this model, the string network is
represented by a collection of uncorrelated straight string segments
moving with uncorrelated, random velocities. There are two fundamental
length scales in such a model: $\xi$, the length of a string segment,
which represents the typical length of roughly straight segments in a
full network; and $\bar{\xi}$, the typical length between two string
segments, which sets the number density of strings in a given volume
($N_s\propto1/\bar{\xi}^2$). The model also tracks the root-mean-square
(rms) velocity of the string segments. These parameters are deduced by
assuming network scaling---i.e., the string energy density tracks the
dominant background energy density in the radiation and matter
eras. Numerical simulations of networks have confirmed that these
assumptions accurately describe network physics. The positions and
orientations of the segments are drawn from uniform distributions in
space and on a two-sphere, respectively. The model's parameters have
been calibrated to produce source correlation functions in agreement
with those in Ref.~\cite{bib:vincent}. Because of the approximations
built into the string segments, the model can only produce two-point
correlation functions. It cannot be used for making simulated string
maps or other more sophisticated observables; though it could, with some
modification, generate three-point functions \cite{Gangui:2001fr}.  

On the cosmological scales we consider, finer details of the string
evolution do not play a major role. It is the large-scale
properties---such as the scaling distance, the equation of state
(wiggliness), and the rms velocity---that determine
the shape of the string-induced spectra. Further details of the
computational methods necessary to compute the 21~cm signal will be
included in a subsequent publication \cite{ustobe}. 

\subsection{Cross correlations}
Let $\delta T^z(\vec{x},\hat{n})$ denote the anisotropy in the 21~cm
brightness temperature emitted at redshift $z$ and observed at a spatial
position $\vec{x}$ coming from the propagation direction
$\hat{n}$. Similarly, the CMB brightness temperature is characterized by
$\delta T^{CMB}(\vec{x}^\prime,\hat{n}^\prime)$. For computational
purposes these quantities are expanded into normal
modes~\cite{Hu:1997hp}  
\be
\delta T^X(\vec{x},\hat{n})=\int
\frac{d^3\vec{k}}{(2\pi)^3}\sum_\ell\sum_{m=-2}^2\delta T_{\ell\,m}^{X}(\vec{k}) \ G_\ell^m\,,
\ee
where $\vec{k}$ is the Fourier conjugate to $\vec{x}$, $X\in\{z,{\rm
CMB}\}$ and $\delta T^X_{\ell\,m}$ are the associated multipole moments
in the basis 
\be
G_\ell^m=(-i)^\ell\sqrt{\frac{4\pi}{2\ell+1}}\,Y_\ell^m(\hat{n})\exp\lb
i\vec{k}\cdot\vec{x}\rb\,.\label{eq:harmonics}
\ee
In our analysis we use the angular correlation function between quantities $X$ and
$Y$ as measured by an observer at $\vec{x}=0$, defined as
\ba
C^{XY}(\Theta)&\equiv& \langle \delta T^z(\hat{n}) \,
\delta T^{CMB}(\hat{n}^\prime)\rangle_{\hat{n}\cdot\hat{n}^\prime=\cos\Theta}\nn\\
&=&\frac{1}{4\pi}\sum_{\ell=0}^\infty(2\ell+1)C_\ell^{XY}\,P_\ell(\cos\Theta)\,
\ea
where $P_\ell$ are the Legendre polynomials, and $C_\ell^{XY}$ is the
anisotropy spectrum calculated from the associated multipoles $\delta
T^X(\vec{k})$~\cite{cmbfast}~\cite{Lewis:2007kz} as 
\be
C_\ell^{XY}=(4\pi)^2\int d^3\vec{k}\,\sum_{m=-2}^2\delta T^X_{\ell\,m}\delta
T^{Y}_{\ell\,m}\,.
\label{eq:xcorr}
\ee
In the above notation, $m=0$ corresponds to scalar modes,  $m=\pm 1$
labels vector modes of right- and left-handed vorticity, and $m=\pm 2$
denotes the tensor modes. See the Appendix for an expanded discussion of
our calculational method.  

\section{Sources of Noise}

Having formed the cross correlation spectrum (Eq.~(\ref{eq:xcorr})), a
density distribution of signal-to-noise $(S/N)$ can be formed in
redshift and $\ell$ space as
\be
\left( \frac{S}{N} \right)^2 (z,\ell,\Delta z)=\frac{f_{\rm sky}(2\ell+1)\left(C_\ell^{Tz}\right)^2}{C_\ell^{TT}C_\ell^{zz}+\lb
C_\ell^{Tz}\rb^2}\,,
\label{eq:s2n}
\ee
$\Delta z$ is the width of the redshift shell set by the chosen frequency
interval and $f_{\rm sky}\simeq0.8$ is the fraction of observable
sky. The total signal-to-noise $(S/N)$ is then 
\be
\lb\frac
SN\rb^2_{\rm tot}=\sum_{\ell}\sum_{z_i}\left( \lb\frac{S}{N}\rb^2 (z_i,\ell,\Delta z_i) \right) \ ,
\ee
where the sum is over the harmonic multipoles $\ell$ and the (independent)
observed redshift slices $z_i$ of width $\Delta z_i$, determined by the
observational setup.
Each multipole can receive contributions from the inflationary
background $C^{ad}_\ell$, the active sources $C^{str}_\ell$, and
astrophysical and telescope noise $C^{N}_\ell$, so we can write
\be
C_\ell^{XY}[\mbox{total}]=C_\ell^{ad\,XY}+C_\ell^{str\,XY}+C_\ell^{N\,XY}\,.
\label{eq:Cl}
\ee

Strings are the only source of the cross-correlation, so the numerator
in Eq.~(\ref{eq:s2n}) receives only one contribution. The denominator,
however, receives a large contribution from inflationary perturbations
that are at least ten times the size of the string power spectrum. There
is also a large contribution from the detector noise which is dominated
by the galactic emission at the low frequencies we need to probe the
high redshift 21~cm brightness temperature. The detector noise is
uncorrelated between the CMB and 21~cm frequencies and is given by  
\be
C_\ell^N={(2\pi)^3 T_{\rm sys}^2(\nu)\over \Delta \nu\, t_{\rm obs} f^2_{\rm
    cover} \ell^2_{\rm max}(\nu)} \ ,
\label{eq:ClTsys}
\ee
where $T_{\rm sys}$ is dominated by the sky temperature and, for regions
away from the galactic plane, can be approximated as
\cite{Furlanetto:2006jb} 
\be
T_{\rm sys}(\nu)=150 {\rm K} \left( \nu \over 180 {\rm MHz}
\right)^{-2.6} \, . 
\ee 
In the previous expressions $\Delta \nu$ is the width of a frequency
bin, which is set by the experiment, $t_{\rm obs}$ is the observing time, 
$f_{\rm cover}\equiv N_{\rm dish}A_{\rm dish}/A_{\rm total}$ is the
fraction of the total area enclosed by the experiment and characterizes
the total collecting area, and $\ell_{max}(\nu)=2\pi D_{\rm array}\nu/c$,
where $D_{\rm array}$ is the diameter of the array and $c$ is the speed
of light.    

Foregrounds include galactic and extragalactic point sources, as well
as the galactic and extragalactic free-free emission, all of which are
expected to induce correlations between different
frequencies~\cite{Zaldarriaga:2003du,Santos:2004ju}. In this study,
however, astrophysical sources will have a negligible effect since we
are only including 21~cm cross-correlations with CMB photons, and these
signals quickly decorrelate with increasing frequency
separation~\cite{Furlanetto:2006jb}. Furthermore, these signals dominate
on very small scales while string-induced
correlations reside on large scales, $\ell\lesssim100$.   

We fold the observational parameters into a single observational
ambition parameter, $x$, defined as 
\be
x\equiv t_{\rm obs} f_{\rm cover}^2 D^2\,,
\ee
with units of $[x]=[years]\times[{\rm km}]^2$. 
There is always a practical trade-off between making $f_{\rm cover}$
large and making $D$ large. For our signal, though, it is clear what we
should choose. The main reason for going to large $D$ is to resolve very
small angular scales. However, string sourced cross correlations reside
only on large angular scales. Hence, we can make due with a relatively
small baseline $D$, which should let us make $f_{\rm cover}$ nearly
unity; i.e., a ``moderately'' sized telescope or array with nearly a
complete covering area would be ideal for searching for strings.

\section{Results}
	
We have calculated cross correlation power spectra from redshifts
throughout the dark ages.  In producing our results, we have fixed the
observing bandwidth to $\Delta\nu=0.5{\rm MHz}$; however,varying this
does not change our result. Although increasing the bandwidth diminishes the
autocorrelator signal (noise ``$N$''$\propto 1/\Delta\nu$) and, thus, increases
the signal-to-noise in each redshift bin, one observes  fewer
independent redshift slices as a result (signal ``$S_{tot}$''$\propto 1/\Delta\nu$)
so the overall signal-to-noise remains unchanged to first order. We have
numerically verified this for bandwidths in the range $0.01{\rm MHz} <
\Delta\nu < 2{\rm MHz}$.    

We have excluded the epoch of reionization ($z<20$) from our
signal-to-noise results. There are several reasons for this. The main
reason is that reionization physics are still too uncertain for reliable
calculations to be done, and the details of reionization, once begun,
will dominate the 21~cm physics. Moreover, if cosmic strings are present
during reionization, they will likely play a role in speeding up the
process of reionization through their early generation of non-linear
overdensities \cite{rees86,Pogosian:2004mi}. Reionization itself will
also create cross-correlations between the CMB and 21~cm
radiation~\cite{Alvarez:2005sa,21cm-xcorr,Adshead:2007ij,Sarkar:2008sz}. Though
it is possible that the string and reionization signals could be
disentangled if they are both present, we restrict ourselves in the
present study to redshifts where strings are the only possible source of
cross-correlation.  

At redshifts $z\lse7$ reionization is complete and the surviving neutral
hydrogen only exist in islands that are presumably correlated with large
scale structure. In principle, we can continue calculating the string
sourced cross correlation between the matter distribution (traced via
21~cm or any other observable light) and the CMB. However, there will be
competing sources of cross correlation. For instance, the re-scattering
of CMB photons off ionized particles will induce a correlation of CMB
with the large-scale plasma distribution
\cite{Giannantonio:2007za}. Eventually, at lower redshifts, there will
be a correlation between large scale structure and the ISW distortion of
the CMB caused time varying potentials when Dark Energy becomes
dynamically important. Nonetheless, we have made an approximate
calculation of what a hypothetical radio survey at these redshifts might
see in the presence of strings, extending an analytical distribution of
large scale structure tracers used in \cite{Hu:2004yd} to the range
$0<z<7$. The resulting theoretical signal-to-noise for the cross
correlation was too small. This is not surprising since string wakes
that are sourced at such late times have not had time to grow. By
contrast, the potential wells that attract the ionized particles which
then re-scatter CMB have been growing through gravitational instability
for some time since entering the horizon. Hence, we do not include the
$z\leq7$ range into our signal-to-noise results.

\begin{figure*}
\subfigure[]
%AB denote $\lb S/N(z,\ell,\Delta z)\rb^2=0.5,\,1.0$.]
{\label{fig:s2n2-stronly}\includegraphics[width=0.56\textwidth]{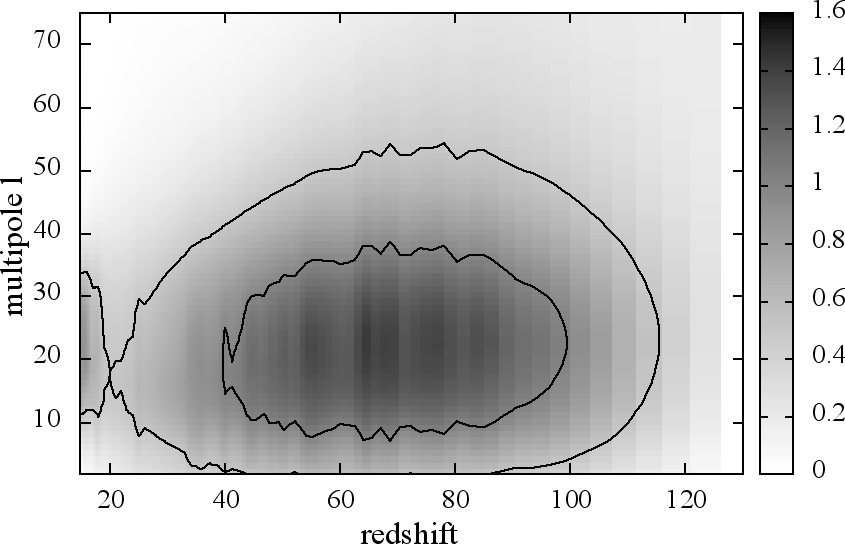}}
 \subfigure[
%AB Contours denote $\lb S/N(z,\ell,\Delta z)\rb^2=0.001,0.002$.]
]{\label{fig:s2n2-inflat}\includegraphics[width=0.56\textwidth]{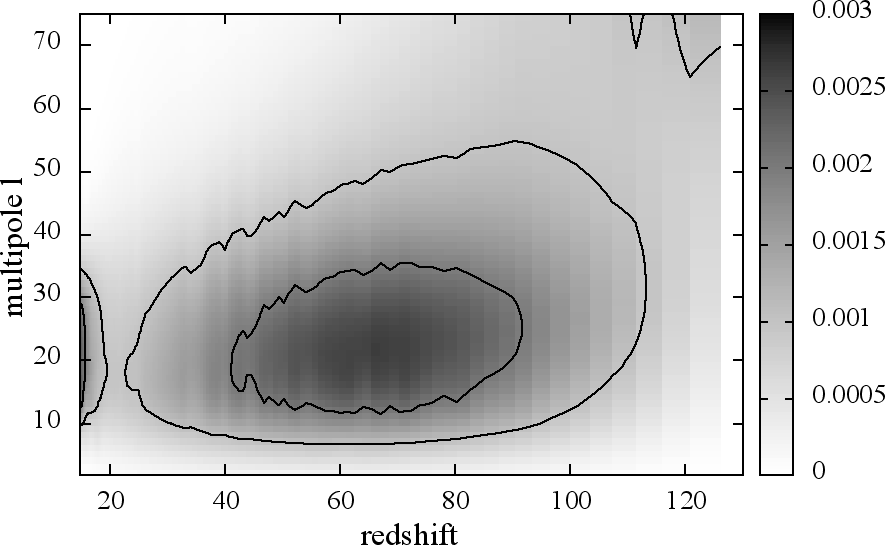}}\quad
\subfigure[
%AB Contours of $\lb S/N(z,\ell,\Delta %z)\rb^2=5\times10^{-6},\,2.5\times10^{-6}$  are shown.]
]
{\label{fig:s2n2-Tsky}\includegraphics[width=0.56\textwidth]{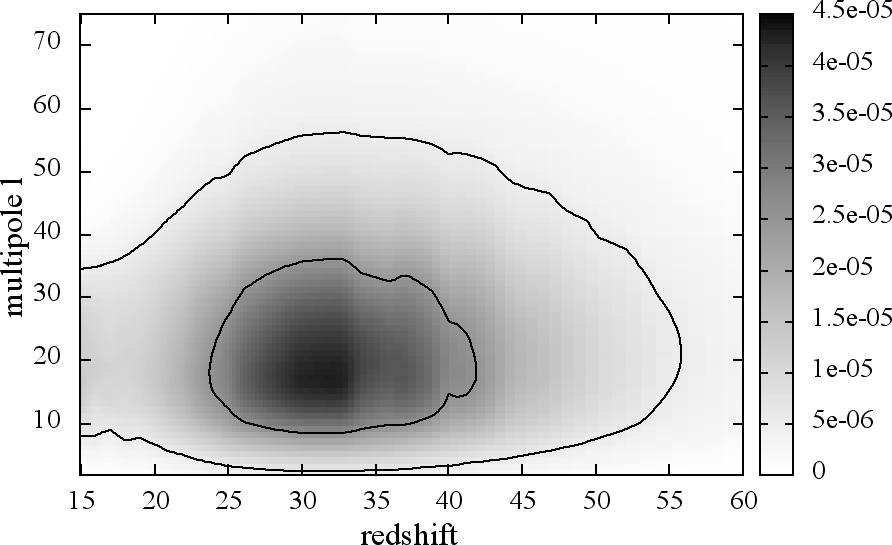}}
\caption{Distribution of signal-to-noise squared $(S/N)^2$ from
 cross-correlation studies between CMB and 21~cm photons in the presence
 of a string network. The observing bandwidth has been fixed to
 $\Delta\nu=0.5{\rm MHz}$, which determines the redshift bin width
 $|\Delta z|=(1+z)^2\Delta \nu/\nu_0$, $\nu_0$ being the rest frequency
 of the 21~cm radiation. Signal along horizontal slices is modulated by
 the brightness temperature, which nulls around the epoch of first light
 as the HI regions transit from absorption to emission. Signal along
 vertical slices is set by the correlation length of the network, a
 fraction of the Hubble scale. Fig.~\ref{fig:s2n2-stronly} shows an
 idealized situation where structure is seeded only by strings, in the
 absence of foreground noise. Contours of $0.5$ and $1.0$ are shown. In
 Figs.~\ref{fig:s2n2-inflat},~\ref{fig:s2n2-Tsky},
 $G\mu=4\times10^{-7}$,  representing a system primarily seeded by
 inflation, but with a 10\% contribution from
 strings. Fig.~\ref{fig:s2n2-inflat} includes no sky noise, and has
 contours representing  $0.001$ and $0.002$. Fig.~\ref{fig:s2n2-Tsky}
 includes noise, and assumes ten year's observation from a square
 kilometre array ($D=1{\rm km}$, $x=10$); recall that $x\equiv t_{\rm
 obs}f_{\rm cover}^2 D^2$. Contours of $5\times10^{-6}$ and
 $2.5\times10^{-6}$ are shown. These plots primarily serve to indicate
 where signal resides in the observing volume. \label{bigfig}}    
\end{figure*}

Let us turn to our results, shown in Fig.~\ref{bigfig}, which begin with
analysis of a hypothetical Universe where inflationary perturbations are
not present and cosmic strings are the only source of both CMB and 21~cm
anisotropy.  We plot the cross correlation signal-to-noise distribution
for this case in  Fig.~\ref{fig:s2n2-stronly}. When only strings are
present and noise is disregarded, as we do here, the overall
normalization, the string tension $G\mu$, drops out of the
signal-to-noise calculation. In this imaginary Universe, the signal is
so strong that an observation at each redshift would generate a
detection ($(S/N)(z,\ell,\Delta z)>1$).  

In the real Universe, the inflationary perturbations overwhelm those
from strings, as we see in Fig.~\ref{fig:s2n2-inflat}, even without
including the detector noise. No single measurement now has significant
signal-to-noise ($(S/N)>1$). Instead, we must sum over many independent
redshift bins in order to accumulate a significant signal, giving a
statistical detection. The case presented in this figure represents the
best-case scenario: no noise and a maximal, ten percent contribution to
primordial anisotropy from strings. The signal is similar in shape to,
though lower in magnitude than, the idealized case of a system seeded
only by strings. The signal is suppressed more significantly on small
angular scales, since the inflationary auto-correlations monotonically
increases for $\ell$, while the string-induced signal peaks at
$\ell\simeq60$.  Although any particular observation in the $z-\ell$
plane will result in a contribution $(S/N)^2\simeq\mathcal{O}(10^{-3})$,
the statistical benefit of observing roughly $50$ multipoles
($2<\ell<50$, weighted by a factor of $2\ell+1$) in roughly $50$
redshift bins (observations with a $1{\rm MHz}$ bandwidth in the range
$20<z<120$) gives rise to a total
$(S/N)\simeq\sqrt{(50)^310^{-3}}\simeq\mathcal{O}(10)$. Summing our
simulated data over all $\ell$ and over $70 \leq z \leq 20$ reveals a
$(S/N)\simeq 3$. However, this does not include experimental noise, and
corresponds to a survey-strategy parameter of $x=\infty$.

Fig.~\ref{fig:s2n2-Tsky} shows the remaining signal once the sky
temperature is incorporated. These data were generated assuming an
all-sky survey from a single dish of diameter $1{\rm km}$ observing for
ten years ($x=10$); note that sky temperature completely overwhelms the
signal for redshift $z>50$. The estimated signal-to-noise remaining in
our analysis volume is $(S/N)\simeq 0.28$.

\section{Observational Outlook}

\begin{figure}[htbp] %  figure placement: here, top, bottom, or page
   \centering
   \includegraphics[width=0.5 \textwidth]{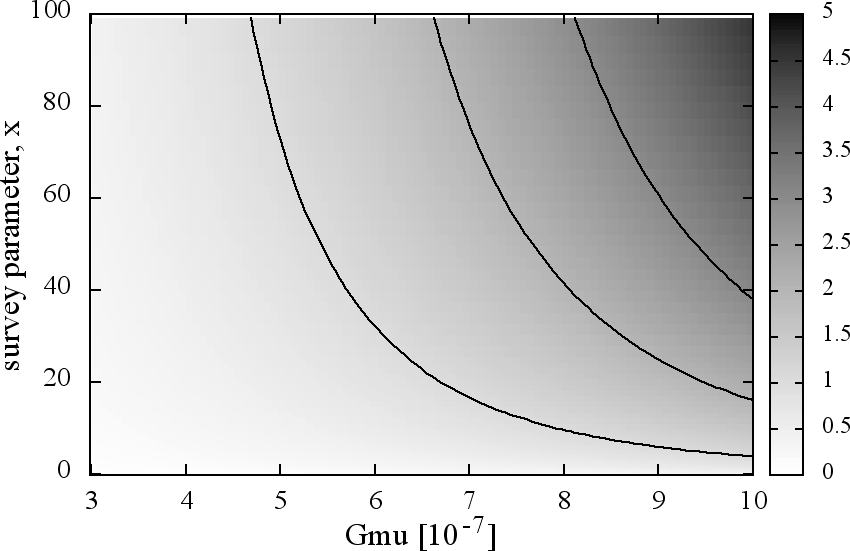} 
   \caption{Signal-to-noise as a function of observing strategy
(represented by $x\equiv t_{\rm obs}f_{\rm cover}^2 D^2$, in units of
 years $\times$ km$^2$) and the string tension $G\mu$. Contours of
 $S/N=1, 2,{\rm and}\, 3$ are shown.\label{fig:s2n-gmu}} 
   \label{fig:example}
\end{figure}

Finally, let us look at Fig.~\ref{fig:s2n-gmu}, which summarizes the
detectability of our signal for various string tensions, $G\mu$, as a
function of our observational parameter, $x$. It is clear that the noise
induced by the large sky temperatures severely limits the possibility of
observing a string network through cross-correlation. However, the
calculation reported here does not exhaust the possible locations of
cross-correlation signal from strings. For instance, our estimate is
restricted to the domain $z>20$ that we are certain will not be affected
by reionization. Our present ignorance about the physics of reionization
means that we cannot trust our calculation into the $z<20$ range.  This
is a major impediment, though: lower $z$ is the redshift range least
affected by noise, since noise goes as an inverse power law with
frequency. Hence, we can hope to boost our signal significantly once the
physics of reionization are better understood. 

We have also only looked at the CMB-21~cm cross-correlations in this
study. Although they are likely very small, we are currently analyzing
contributions from the $\mathcal{O}(10^3)$ 21~cm-21~cm
cross-correlations. These will also give some boost to our signal.

\section{Discussion}

By incorporating the segment model of strings into the CAMB\_sources
Einstein-Boltzmann code, we have calculated
the cross correlation between the 21~cm brightness
temperature and the CMB. Our chief result is plotted
in Fig.~\ref{fig:s2n-gmu}, which represents the signal-to-noise
available in CMB-21~cm correlation studies given an all-sky survey, a
string tension $G\mu$, and our observing strategy, which we characterize
by a single integration time and telescope properties parameter,
$x$. Unfortunately, noise from the sky temperature overwhelms the signal
we are calculating for observationally-allowed cosmic string tensions. 

However, there is room for more work. We have thus far only included the
signal from the CMB-21~cm correlations for $z>20$, where we do not need
to know the physics of reionization. Since the signal should extend
until the epoch of reionization $z\simeq10$, we may eventually be able
to include  $\mathcal{O}(10^2)$ more redshift bins and
$\mathcal{O}(10^3)$ data points in the $z-\ell$ plane, once we can
accurately model reionization.  Another contribution to our signal that
was neglected in this study can be found from a cross-correlation study
between different redshift bins, and this signal will be addressed in
future work~\cite{ustobe}. It is possible that we will find a detectable
signal when these extra sources of data are included in the analysis. 

Though our calculations have not predicted a detectable signal, we
reiterate that any detection of a cross correlation between the 21~cm
radiation from $z>20$ and the CMB would be a clear sign of new
physics. Active sources, such as cosmic strings, are a promising
candidate for the sort of new physics that could generate this
correlation. However, the fact that inflation very likely provides the
dominant source of perturbations in the Universe makes detectability of
any signal difficult: the inflationary perturbations must necessarily
enter as ``noise'' for any study like this, and are typically much
larger than any alternate sources of perturbation, at least prior to
recombination where measurements of the CMB limit any non-inflationary
perturbations. One way around this would be to contrive some active
source that was suppressed prior to recombination which subsequently
sourced a larger fraction of perturbations in the dark ages. On the
other hand, a non-detection of such a cross correlation would constitute
a strong constraint on active sources with such unusual properties.

\acknowledgements

A.B. is supported by the National Sciences and Engineering Research
Council of Canada (NSERC) National Postdoctoral Fellowship. L.P. is
supported by an NSERC Discovery grant. The work of M.W. at the Perimeter
Institute is supported by the Government of Canada through Industry
Canada and by the Province of Ontario through the Ministry of Research
\& Innovation.

%\begin{figure}
%\includegraphics[width=.49\textwidth]{./s2n2-matter.png}
%\label{fig:s2n2-matter}
%\caption{The distribution of signal to noise (squared) for
% string-induced matter-CMB correlations. Signal is present at two
% distinct scales, and can be attributed to the correlation scales of the
% string network.}
%\end{figure}

\appendix

\section{21~cm vector modes}

In the absence of active sources vector perturbations quickly decay and,
for that reason, are frequently ignored.  In particular, they are not
included in the 21~cm extension to CAMB~\cite{Lewis:2007kz}. Cosmic
strings, however, actively source vector modes and provide a significant
contribution to CMB and 21~cm anisotropies. A large effort was spent
incorporating their effects into our numerical code~\cite{ustobe}, and
we elaborate on some of the details here.

Applying the harmonic decomposition described in~\cite{Hu:1997hp} and
Eq.~(\ref{eq:harmonics}) to (fluctuations in) the photon distribution
function $f$ provided in~\cite{Lewis:2007kz} gives scalar $m=0$, vector
$m=\pm 1$ and tensor $m=\pm 2$ contributions to the expanded field:
\ba
\delta f(\eta,\vec{x},\epsilon,\hat{n})=\int
\frac{d^3\vec{k}}{(2\pi)^3}\sum_{l,m} F_l^{(m)}(\eta,\epsilon,\vec{k})\,{_0{G_l^m}}\,.
\label{eq:harm.expansion}
\ea
The photon fluid is characterized at some conformal time $\eta$ and
position $\vec{x}$ as photons coming from some direction $\hat{n}$ with
an energy $\epsilon=a(\eta)E_{21}$. The quantity $\epsilon$ represents
the Doppler shifted energy of a 21~cm photon, $E_{21}$, stretched by the
scale factor $a(\eta)$. $\vec{k}$ is the Fourier conjugate to the
spatial coordinate $\vec{x}$, we denote $k=|\vec{k}|$. 

Expressions for the harmonic multipoles $F_l^{(m)}$ are formally obtained by
integrating the Boltzmann equation for 21~cm photons along the line of
sight~\cite{cmbfast}. The scalar contribution $F_l^{(0)}$ has been reported
in~\cite{Lewis:2007kz}, while for vector modes we find
\begin{widetext}
\ba
\hspace{-1pt}F_l^{(1)}(\eta_A,\epsilon,\vec{k})& \simeq &
\hspace{-4pt}
-e^{-\tau_c}\bar{f}(\epsilon)\frac{r_\tau}{\mathcal{H}}v^{(1)}\lb\frac{j_l^\prime(k\chi_\epsilon)}{\chi_\epsilon}-\frac{j_l(k\chi_\epsilon)}{k\chi_\epsilon^2}\rb+
\int_{\eta_\epsilon}^{\eta_A}\hspace{-8pt}d\eta \;
 \dot\tau_c \,e^{-\tau_c}\bar{f}_{,\ln\epsilon}(\epsilon)v^{(1)}\,\frac{j_l(k\chi)}{k\chi}
\\
&& - e^{-\tau_\epsilon}\lc\bar{f}(\epsilon)\cdot\ls
v^{(1)}+\frac{\bar{T}_\gamma}{\bar{T}_s-\bar{T}_\gamma}(v^{(1)}_\gamma-v^{(1)})+\frac{r_\tau}{\mathcal{H}}\dot
v^{(1)}+(r_\tau-1)v^{(1)}\rs+\bar{f}_{,\ln\epsilon}(\epsilon)v^{(1)}\rc_\epsilon\hspace{-2.5pt}\frac{j_l(k\chi_\epsilon)}{k\chi_\epsilon}
\nn
\\
&&-e^{-\tau_c}\bar{f}(\epsilon)\frac{\bar{T}_\gamma}{\bar{T}_s-\bar{T}_\gamma}\sum_{l^\prime=2}^\infty
%%AB: subbed in analog. expression\Theta_{l^\prime} j_l^{(l^\prime 1)}(k\chi_\epsilon)
\Theta_{l^\prime}P_{l^\prime}\lb\frac{-i}{k}\frac{d}{d\chi_\epsilon}\rb j_l(k\chi_\epsilon)
+\int_{\eta_\epsilon}^{\eta_A}d\eta\;\dot\tau_ce^{-\tau_c}\bar{f}_{,\ln\epsilon}(\epsilon)\frac{\sqrt{3}\,F_2}{4}\lb\frac{j_l^\prime(k\chi_\epsilon)}{\chi_\epsilon}-\frac{j_l(k\chi_\epsilon)}{k\chi_\epsilon^2}\rb\nn
\label{vec}
\ea
\end{widetext}
for $l\ge1$, where a prime denotes a derivatives with respect to the
argument and a bar denotes a background value. Here, $\tau_c$ is the
Thomson scattering optical depth, $\tau_\epsilon$ is the optical depth
to 21~cm radiation, $v^{(1)}$ is the vector component to the baryon
velocity, $v_\gamma^{(1)}$ is the velocity (dipole) of the photon
distribution, $\bar{f}$ is the background photon distribution function,
$\bar{T}_\gamma$ is the photon temperature, $\bar{T}_s$ is the spin-flip
temperature of the neutral hydrogen and determines how much the hydrogen
is emitting,  
$r_\tau \equiv \tau_\epsilon e^{-\tau_\epsilon}/(1-e^{-\tau_\epsilon})$,
$\mathcal{H}$ is the conformal Hubble parameter,
$\chi_\epsilon\equiv\eta_A -\eta_\epsilon$ is the conformal distance
along the line of sight, and $j_l(k\chi_\epsilon)$ is the  spherical
Bessel function. Quantities subscripted with $\epsilon$ are evaluated at
the conformal time $\eta_\epsilon$.  The $\Theta_\ell$ are the angular
moments of the Fourier expansion of the CMB temperature anisotropy. All
quantities are discussed at length in~\cite{Lewis:2007kz}. 

The 21~cm power spectra referenced in the main text were given in terms
of the brightness temperature $T_b$. This relates to the photon distribution
function $f$ through the relation
\be
T_b=\frac{E_{\rm obs}h_p^3 f}{2k_b}
\ee
where $E_{\rm obs}$ is the observed frequency of the 21~cm emission,
$h_p$ is the Plank constant, and $k_b$ is the Boltzmann constant. Thus,
anisotropies in the 21~cm brightness temperature emitted at a redshift
$z$ are given as
\be
\delta T^z(\vec{x},\hat{n})=\frac{E_{\rm obs}h_p^3 }{2k_b}\,\delta
f(\eta,\vec{x},\epsilon,\hat{n})\,. 
\ee

\end{document}